\documentclass[12pt]{iopart}
\usepackage{cite} 
\usepackage{wrapfig}
\usepackage{graphicx}
\usepackage{graphics}

\begin{document}

\title[A bosonic multi-state two-well model]{A bosonic multi-state two-well model}

\author{ G. Santos$^{1}$, A. Foerster$^{2}$ and I. Roditi$^{1}$}

\address{$^{1}$ Centro Brasileiro de Pesquisas F\'{\i}sicas - CBPF \\ 
Rua Dr. Xavier Siguad, 150, Urca, Rio de Janeiro - RJ - Brazil}
\ead{gfilho@cbpf.br,roditi@cbpf.br}
\address{$^{2}$ Instituto de F\'{\i}sica da UFRGS \\ 
Av. Bento Gon\c{c}alves, 9500, Agronomia, Porto Alegre - RS - Brazil}
\ead{angela@if.ufrgs.br}


\begin{abstract}
Inspired by the increasing possibility of experimental control in ultra-cold atomic physics we introduce a new Lax operator and use it to construct and solve models with two wells and two on well states together with its generalization for $n$ on well states. The models are solved by the algebraic Bethe ansatz method and can be viewed as describing two Bose-Einstein condensates allowing for an exchange interaction through Josephson tunneling.
\end{abstract}

\section{Introduction}

The realization of Bose-Einstein condensates (BEC), achieved by taking dilute alkali gases to ultra low temperatures \cite{early,angly} 
is certainly among the most exciting recent experimental achievements in physics. Since then, investigations dedicated to the comprehension 
of  new phenomena associated to this state of matter as well as its properties have flourished, either in the experimental or theoretical domains. 
One noticeable recent effort is the one that proposes the study of a two-well model with two levels at each well as a mean to study Einstein-Podolsky-Rosen 
entanglement \cite{drummond}.

This quest encouraged the search for new solvable models that could be related to the properties of such condensates, including the possibility 
of interaction among condensates  \cite{jon1,jon2,jonjpa,key-3,dukelskyy,Ortiz,Kundu,eric5,GSantosaa,GSantos,GSantos11}. The motivation that underlies 
those proposed models is that, by the study of exactly solvable models, quantum fluctuations may be fully taken into account providing tools 
that allows one to go beyond the results obtained by mean field approximations. We believe that this fruitful approach may furnish some new insights in 
this area, and contribute as well to the increasingly interesting field of integrable systems itself \cite{hertier, batchelor}. In the present paper, we will 
use the algebraic Bethe ansatz method to obtain a new multilevel two-well model. Each well is linked to a BEC and tunneling between the levels of each well is allowed. 
The algebraic formulation of the Bethe ansatz, and the associated quantum inverse scattering method (QISM), was primarily  developed in\cite{fst,ks,takhtajan,korepin,faddeev}.

The QISM has been used to unveil properties of a considerable number of  solvable systems, such as, one-dimensional spin chains, quantum field theory of one-dimensional 
interacting bosons \cite{korepin1} and fermions \cite{yang}, two-dimensional lattice models \cite{korepin2}, systems of strongly correlated electrons \cite{ek,ek2}, conformal field theory \cite{blz}, integrable systems in high energy physics \cite{lipatov, korch, belitsky}
 and quantum algebras (deformations of universal
enveloping algebras of Lie algebras) \cite{jimbo85,jimbo86,drinfeld,frt}.  For a pedagogical and historical review see \cite{faddeev2}. More recently solvable 
models have also showed up in relation to string theories (see for instance \cite{dorey}).  Remarkably it is important to mention that exactly solvable models 
are recently finding their way into the lab, mainly in the context of ultracold atoms \cite{batchelor2} but also in nuclear magnetic resonance (NMR) experiments\cite{kino1,kino2,kitagawa,haller,liao,coldea,nmr1}
turning its study as well as the derivation of 
new models an even more fascinating field.

Our point of view here comes from very recent results concerning the construction of Lax operators, by which it is possible to obtain solvable models suitable for 
the effective description of the interconversion interactions occurring in the BEC. Acquiring our motivation from
these ideas we present the construction of a two-well solvable model that contemplates interconversion among the levels in each well. We obtain this model by a 
multistate Lax operator whose construction is fully explained in the sequel. It is motivated by the construction in \cite{Kuznet, key-3}, where a Lax operator is 
defined for a single canonical boson operator, but instead of a single operator we choose a linear combination of independent canonical boson operators.

It is convenient to underline that although in the sequel we follow a formal presentation we do arrive at new integrable physical Hamiltonians that share many of the aspects of the physical systems studied through interferometric techniques, such as in \cite{QYHEDrumm,kuhnert}. In particular, as our models are solvable, one can obtain precise results, for instance, of properties related to the energy gap, entanglement and ground state fidelity \cite{rubeni}. Also, as mentioned above, increasingly sophisticated NMR techniques \cite{oliv, nmr1} allows manipulation of qubits and we believe that with our models we add an interesting possibility to the usual NMR nuclear quadrupole Hamiltonian.

\section{Algebraic Bethe ansatz method}

In this section we will shortly review the algebraic Bethe ansatz method and present the transfer matrix used to get the solution of the 
models \cite{jonjpa,Roditi}. We begin with the $gl(2)$-invariant $R$-matrix, depending on the spectral parameter $u$,

\begin{equation}
R(u)= \left( \begin{array}{cccc}
1 & 0 & 0 & 0\\
0 & b(u) & c(u) & 0\\
0 & c(u) & b(u) & 0\\
0 & 0 & 0 & 1\end{array}\right),\end{equation}

\noindent with $b(u)=u/(u+\eta)$, $c(u)=\eta/(u+\eta)$ and $b(u) + c(u) = 1$. Above,
$\eta$ is an arbitrary parameter, to be chosen later.

It is easy to check that $R(u)$ satisfies the Yang-Baxter equation

\begin{equation}
R_{12}(u-v)R_{13}(u)R_{23}(v)=R_{23}(v)R_{13}(u)R_{12}(u-v)
\end{equation}

\noindent where $R_{jk}(u)$ denotes the matrix acting non-trivially
on the $j$-th and the $k$-th spaces and as the identity on the remaining
space.

Next we define the monodromy matrix  $T(u)$,

\begin{equation}
T(u)= \left( \begin{array}{cc}
 A(u) & B(u)\\
 C(u) & D(u)\end{array}\right),\label{monod}
\end{equation}

\noindent such that the Yang-Baxter algebra is satisfied

\begin{equation}
R_{12}(u-v)T_{1}(u)T_{2}(v)=T_{2}(v)T_{1}(u)R_{12}(u-v).\label{RTT}
\end{equation}

\noindent In what follows we will choose a realization for the monodromy matrix $\pi(T(u))=L(u)$  
to obtain solutions of a family of models for multilevel two-well Bose-Einstein condensates.
In this construction, the Lax operators $L(u)$  have to satisfy the relation

\begin{equation}
R_{12}(u-v)L_{1}(u)L_{2}(v)=L_{2}(v)L_{1}(u)R_{12}(u-v),
\label{RLL}
\end{equation}

Then, defining the transfer matrix, as usual, through

\begin{equation}
t(u)= tr \;\pi (T(u)) = \pi(A(u)+D(u)),
\label{trTu}
\end{equation}
\noindent it follows from (\ref{RTT}) that the transfer matrix commutes for
different values of the spectral parameter; i. e.,

\begin{equation}
[t(u),t(v)]=0, \;\;\;\;\;\;\; \forall \;u,\;v.
\end{equation}
\noindent Consequently, the models derived from this transfer matrix will be integrable. Another consequence is that the 
coefficients $\mathcal{C}_k$ in the transfer matrix $t(u)$,

\begin{equation}
t(u) = \sum_{k} \mathcal{C}_k u^k,
\end{equation}
\noindent are conserved quantities or simply $c$-numbers, with

\begin{equation}
[\mathcal{C}_j,\mathcal{C}_k] = 0, \;\;\;\;\;\;\; \forall \;j,\;k.
\end{equation}

If the transfer matrix $t(u)$ is a polynomial function in $u$, with $k \geq 0$, it is easy to see that,

\begin{equation}
\mathcal{C}_0 = t(0) \;\;\; \mbox{and} \;\;\; \mathcal{C}_k = \frac{1}{k!}\left.\frac{d^kt(u)}{du^k}\right|_{u=0}. 
\label{C14b}
\end{equation}

\section{Multi-State Lax Operators}

 Now we introduce a new $L$ operator with multi-state bosonic components. We have $n$ operators $\hat O^{r}_{j}$ each one acting on a given state, here the index $j$ means the state and the index $r$ means the site corresponding to the Lax operator. In our case we will consider only two sites, each one supposed to be a well containing a Bose-Einstein condensate. In other words the operators act, for each site, on the direct sum of the spaces associated to those states,

\begin{equation}
\label{eq.1}
 V = V_1 \oplus V_2 \oplus ... \oplus V_n.
\end{equation}

The operators of different states or different sites commute, 

\begin{equation}
 [\hat{O}^r_{j},\hat{O}^s_{k}] = 0 \, \forall \, r \neq s \,\,\mathrm{and}\,\, j\neq k,
\end{equation}
\noindent and for the same state and site they obey their respective algebras. More explicitly, for the usual bosonic operators satisfying the canonical commutation relations ( and from hereafter we drop the $r$ index as we denote one site $a$ and the other site $b$),

\begin{equation}
 [a_{i}^{\dagger},a_{j}^{\dagger}] = [a_{i},a_{j}] = 0,\;\;\;[a_{i},a_{j}^{\dagger}]=\delta_{ij}I.
\end{equation}
\noindent we have the following solution for a multi-state Lax operator,

\begin{equation}
L^{\Sigma_{a}}(u) = \left(\begin{array}{cc}
uI + \eta\sum_{j=1}^{n}N_{aj} & \sum_{j=1}^{n}t_{j}a_{j}\\
\sum_{j=1}^{n}s_{j}a_{j}^{\dagger} & \eta^{-1}\zeta I
\end{array}\right),
\label{L2}
\end{equation}

\noindent if the condition, $\zeta = \sum_{j=1}^{n}s_{j}t_{j}$, is satisfied, where $\zeta$ is a constant value. The above Lax operator satisfies then equation (\ref{RLL}).

Viewed as a monodromy matrix (\ref{monod}) the Lax operator (\ref{L2}) has the following  identifications

$$
A(u) = uI + \eta\sum_{j=1}^{n}N_{aj}, \qquad B(u) = \sum_{j=1}^{n}t_{j}a_{j},$$

$$
C(u) = \sum_{j=1}^{n}s_{j}a_{j}^{\dagger}, \qquad D(u) = \eta^{-1}\zeta I,
$$
\noindent and the commutation relations,

$$
[A(u),B(v)] = - \eta B(v),  \qquad  [A(u),C(v)] = \eta C(v) ,$$

$$
[B(u),C(v)] = \zeta I,  \qquad [\star, D(u)] = 0, 
$$
\noindent where $\star$ stand for $A(v),\;B(v),C(v)$ or $D(v)$.

\section{Models}

In this section we present two applications of the Lax operator $L$ in (\ref{L2}). The two-well model for two on well states and its generalization for $n$ on well states.

\subsection{The two-well model with two on well states}

The Hamiltonian of the system for two wells (sites)  $a$ and $b$ is,

\begin{eqnarray}
	H &=& U_{aa11}N_{a1}^2 + U_{aa12}N_{a1}N_{a2} + U_{aa22}N_{a2}^2 \nonumber \\ 
	&+& U_{bb11}N_{b1}^2 + U_{bb12}N_{b1}N_{b2} + U_{bb22}N_{b2}^2  \nonumber \\
	&+& U_{ab11}N_{a1}N_{b1} + U_{ab12}N_{a1}N_{b2} \nonumber \\
	&+& U_{ab21}N_{a2}N_{b1} + U_{ab22}N_{a2}N_{b2} \nonumber \\
	&-&  \mu_{1}(N_{a1} - N_{b1}) - \mu_{2}(N_{a2} - N_{b2}) \nonumber \\
	&+&  \epsilon_{a1} N_{a1} + \epsilon_{a2} N_{a2} + \epsilon_{b1} N_{b1} + \epsilon_{b2} N_{b2}\nonumber \\
	  &-& \Omega_{11}(a_{1}^{\dagger}b_{1} + b_{1}^{\dagger}a_{1}) - \Omega_{12}(a_{1}^{\dagger}b_{2} + b_{2}^{\dagger}a_{1}) \nonumber \\ 
	  &-&  \Omega_{21}(a_{2}^{\dagger}b_{1} + b_{1}^{\dagger}a_{2}) - \Omega_{22}(a_{2}^{\dagger}b_{2} + b_{2}^{\dagger}a_{2}).
\label{H1}
\end{eqnarray}
\noindent In the diagonal part of the Hamiltonian (\ref{H1}),  the $U_{pqjk}$ parameters describe the atom-atom $S$-wave scattering in the wells, the $\mu_{j}$ parameters are the relative external potentials between the wells for the on well states $\epsilon_{pj}$. The operators $N_{pj}$ are the number of atoms operators. The labels $p$ and $q$ stand for the wells $a$ and $b$, and the labels $j$ and $k$ stand for the on well states $1$ and $2$.  In the off diagonal part of the Hamiltonian the parameters $\Omega_{jk}$ are the tunnelling amplitudes. 

In the Fig. 1 we show a two-well potential with their respective on well states, where $\epsilon_{a1}$, $\epsilon_{a2}$ are the two states in the well $a$ and $\epsilon_{b1}$, $\epsilon_{b2}$  are the two states in the well $b$. The external potentials, $\mu_j$, shift their respective on well states.

\begin{figure}
\begin{center}
\includegraphics[scale=0.3]{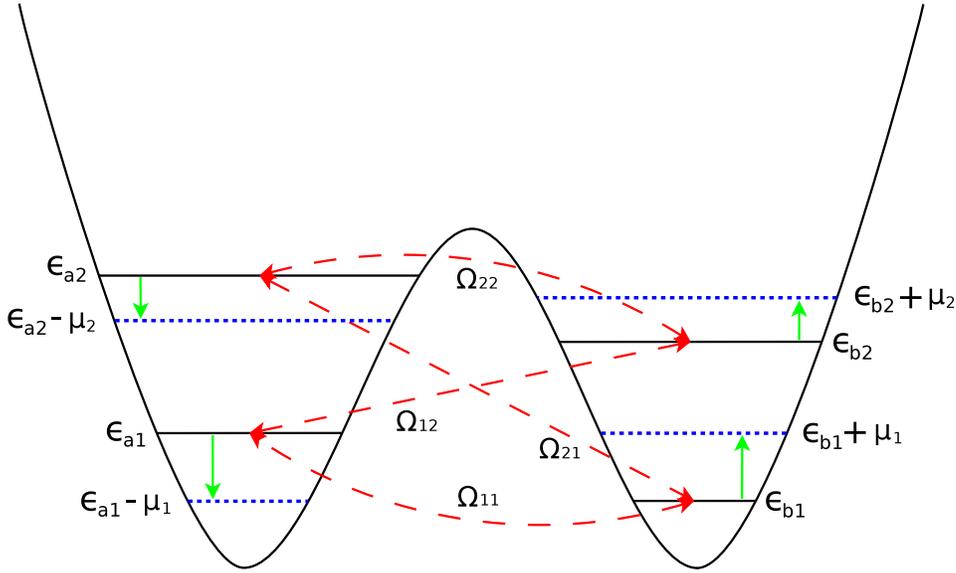}
\caption{(Color online) In the figure we have two wells coupled by Josephson tunnelling with two on well states in each well. The atoms can tunnel between the wells for the same or different states. The dashed red arrows  and the respective tunnelling amplitudes, $\Omega_{jk}$, between the wells for the same or different state are shown. We have two external potentials, $\mu_1$ and $\mu_2$, that shift the on well states. The dotted blue lines and the green arrows show two possible shifts, with $\mu_1 \neq \mu_2$. }
\end{center}
\label{f1}
\end{figure}

It is important to note that if we turn off the tunnelling,\;$\Omega_{jk}=0$, the Hamiltonian (\ref{H1}) describes two decoupled Bose-Einstein condensates with two states in each one, $\{|n_{pj}\rangle\},\;j=1,2,\;p=a,b,$  and only one pure vector state for each condensate $|\psi_p\rangle = |n_{p1}\rangle\otimes|n_{p2}\rangle,\;p=a,b$, with the total vector state of the system the tensor product of those pure vectors states,

 $$|\Psi_T\rangle = |n_{a1}\rangle\otimes|n_{a2}\rangle\otimes|n_{b1}\rangle\otimes|n_{b2}\rangle.$$ 

The energies, $E_a$ and $E_b$, are respectively,

\begin{eqnarray}
	E_a &=& U_{aa11}n_{a1}^2 + U_{aa12}n_{a1}n_{a2} + U_{aa22}n_{a2}^2 \nonumber \\
	  &+&  (\epsilon_{a1} - \mu_{1})n_{a1} +  (\epsilon_{a2} - \mu_{2})n_{a2}, 
\label{Ea}
\end{eqnarray}

\begin{eqnarray}
	E_b &=& U_{bb11}n_{b1}^2 + U_{bb12}n_{b1}n_{b2} + U_{bb22}n_{b2}^2 \nonumber \\
	  &+&  (\epsilon_{b1} + \mu_{1})n_{b1} +  (\epsilon_{b2} + \mu_{2})n_{b2}.
\label{Eb}
\end{eqnarray}
\noindent The total energy is, $E = E_a + E_b$, and the ground state in each condensate depends only of the scattering interactions $U_{aaij}$ and $U_{bbij}$, and the external potentials $\mu_j$. We have four conserved quantities, $[H,I_1]=[H,I_2]=[H,I_3]=[H,I_4] = 0$, with $I_1 = N_{a1},\; I_2 = N_{a2},\;I_3 = N_{b1}$ and $I_4 = N_{b2}$, and so the total number of atoms is a conserved quantity, $N =  I_1 + I_2 + I_3 + I_4$.

When we turn on the tunnelling,\;$\Omega_{jk} \neq 0$,the BEC are now coupled by Josephson tunnelling and the total number of atoms,

$$N = N_{a1} + N_{a2}  + N_{b1} + N_{b2},$$

\noindent continues a conserved quantity, but now, $I_j,\; j = 1,2,3,4$, are no more conserved because of the tunnelling amplitudes,

$$[H,N] = 0,\qquad [H,I_j] \neq 0.$$
\noindent  If $\Omega_{12}=\Omega_{21}=0$, we have another conserved quantities, $[H,J_1]=[H,J_2]=0$, with $J_1 = N_{a1} + N_{b1}$ and $J_2 = N_{a2} + N_{b2}$, and so the total number of atoms is a conserved quantity, $N =  J_1 + J_2$.

 The state space is spanned by the base $\{|n_{a1},n_{a2},n_{b1},n_{b2}\rangle\}$ and we can write each vector state as 

\begin{equation}
  |n_{a1},n_{a2},n_{b1},n_{b2}\rangle = \frac{1}{\sqrt{n_{a1}!n_{a2}!n_{b1}!n_{b2}!}}(a_{1}^{\dagger})^{n_{a1}}(a_{2}^{\dagger})^{n_{a2}}(b_{1}^{\dagger})^{n_{b1}}(b_{2}^{\dagger})^{n_{b2}}|0\rangle,
\label{state1}   
\end{equation}

\noindent where $|0\rangle = |0_{a1},0_{a2},0_{b1},0_{b2}\rangle$ is the vacuum vector state in the Fock space. We can use the states (\ref{state1}) to write the matrix representation of the Hamiltonian (\ref{H1}). The dimension of the space increase very fast when we increase $N$,

$$	d = \frac{1}{6}(N + 3)(N + 2)(N + 1), $$

\noindent with $N$ as a constant $c$-number, $N = n_{a1} + n_{a2} + n_{b1} + n_{b2}$.

Now we use the co-multiplication property of the Lax operators to write the following monodromy matrix,

\begin{equation}
  L(u) = L_{1}^{\Sigma a}(u + \sum_{j=1}^{2}\omega_{j})L_{2}^{\Sigma b}(u - \sum_{j=1}^{2}\omega_{j}).
\label{LH1}
\end{equation}
\noindent Following the monodromy matrix (\ref{monod}) we can write the operators,

\begin{eqnarray}
\pi(A(u)) &=& [(u + \omega_1 + \omega_2)I + \eta\sum_{j=1}^2 N_{aj}][(u  - \omega_1 - \omega_2)I + \eta\sum_{j=1}^2 N_{bj}] \nonumber \\
          &+& \sum_{j,k=1}^2 s_jt_kb_j^{\dagger}a_k, \label{piA}\\
\pi(B(u)) &=& [(u + \omega_1 + \omega_2)I + \eta\sum_{j=1}^2 N_{aj}]\sum_{j=1}^2 t_jb_j + \eta^{-1}\zeta\sum_{j=1}^2 t_jb_j, \label{piB}\\
\pi(C(u)) &=& (\sum_{j=1}^2 s_ja_j^{\dagger})[(u  - \omega_1 - \omega_2)I  + \eta\sum_{j=1}^2 N_{bj}] \nonumber \\
          &+& \eta^{-1}\zeta\sum_{j=1}^2 t_jb_j^{\dagger}, \label{piC}\\
\pi(D(u)) &=& \sum_{j,k=1}^2 s_jt_ka_j^{\dagger}b_k + \eta^{-2}\zeta^2 I. \label{piD}
\end{eqnarray}

Taking the trace of the operator (\ref{LH1}) we get the transfer matrix

\begin{eqnarray}
t(u) & = & u^2I + u\eta N + [\eta^{-2}\zeta^{2} - (\omega_{1} + \omega_{2})^2]I  \nonumber\\
 & + & \eta^{2}(N_{a1}N_{b1}+N_{a1}N_{b2}+N_{a2}N_{b1}+N_{a2}N_{b2}) \nonumber \\
  & + & \eta(\omega_{1} + \omega_{2})(N_{b1} - N_{a1} + N_{b2} - N_{a2}) \nonumber\\
 & + & s_{1}t_{1}(a_{1}^{\dagger}b_{1} + b_{1}^{\dagger}a_{1}) + s_{1}t_{2}(a_{1}^{\dagger}b_{2} + b_{2}^{\dagger}a_{1}) \nonumber\\
 & + & s_{2}t_{1}(a_{2}^{\dagger}b_{1} + b_{1}^{\dagger}a_{2}) + s_{2}t_{2}(a_{2}^{\dagger}b_{2} + b_{2}^{\dagger}a_{2}).
\label{tH1} 
\end{eqnarray}
\noindent From (\ref{C14b}) we identify the conserved quantities of the transfer matrix,

\begin{eqnarray}
	\mathcal{C}_0 &=& [\eta^{-2}\zeta^{2} - (\omega_{1} + \omega_{2})^2]I \nonumber\\
 & + & \eta^{2}(N_{a1}N_{b1}+N_{a1}N_{b2}+N_{a2}N_{b1}+N_{a2}N_{b2}) \nonumber \\
  & + & \eta(\omega_{1} + \omega_{2}) (N_{b1} - N_{a1} + N_{b2} - N_{a2}) \nonumber\\
 & + & s_{1}t_{1}(a_{1}^{\dagger}b_{1} + b_{1}^{\dagger}a_{1}) + s_{1}t_{2}(a_{1}^{\dagger}b_{2} + b_{2}^{\dagger}a_{1}) \nonumber\\
 & + & s_{2}t_{1}(a_{2}^{\dagger}b_{1} + b_{1}^{\dagger}a_{2}) + s_{2}t_{2}(a_{2}^{\dagger}b_{2} + b_{2}^{\dagger}a_{2}),
\end{eqnarray} 
 
\begin{eqnarray}
 \mathcal{C}_1 &=&  \eta N,
\end{eqnarray}

\begin{eqnarray}
 \mathcal{C}_2 &=& I. 
\end{eqnarray}

The Hamiltonian (\ref{H1}) is related with the transfer matrix (\ref{tH1}) by the equation,

\begin{equation}
 H = u^2I + u\mathcal{C}_1  + \frac{\alpha}{\eta^2}\mathcal{C}_1^2   + [\eta^{-2}\zeta^{2} - (\omega_{1} + \omega_{2})^2]I - t(u),
\label{H2} 
\end{equation}
\noindent where we have the following identification between the parameters,

$$\alpha =  U_{aajj} = U_{bbjj}, \qquad 2\alpha = U_{aajk} = U_{bbjk} \; (j\neq k),$$

$$ 2\alpha - \eta^2 = U_{abjk},$$

$$ \eta(u - \omega_{1} - \omega_{2}) = \epsilon_{aj} - \mu_{j},\qquad  \eta(u + \omega_{1} + \omega_{2}) = \epsilon_{bj} + \mu_{j} $$

$$ \Omega_{jk} = s_j t_k, \qquad \sum_{j=1}^2 s_j t_j = \zeta, \qquad j,k = 1,2. $$

\noindent Using as pseudo-vacuum the tensor product vector state, $|0 \rangle \equiv |0,0\rangle_a \otimes |0,0 \rangle_b$, with $|0,0 \rangle_p$, which belongs to the direct sum space, denoting the Fock vacuum state associated to the well $p$ $(p = a,b)$, we can apply the algebraic Bethe ansatz method in order to find the Bethe ansatz equations (BAE),

\begin{eqnarray}
\frac{\eta^{2}[v^2_{i} - (\omega_{1} + \omega_{2})^2]}{\zeta^2} & = & \prod_{j\ne i}^{N}\frac{v_{i}-v_{j}-\eta}{v_{i}-v_{j}+\eta}, \;\;\;\;\;
 i,j=1,\ldots , N.
\label{BAE1}
\end{eqnarray}

The eigenvectors $\{ |v_1,v_2,\ldots,v_N\rangle \}$ of the Hamiltonian (\ref{H1}) or (\ref{H2}) and of the transfer matrix (\ref{tH1}) are

\begin{equation}
|\vec{v}\rangle \equiv  |v_1,v_2,\ldots,v_N\rangle = \prod_{i=1}^N \pi(C(v_i))|0 \rangle,
\end{equation}

\noindent and the eigenvalues of the Hamiltonian (\ref{H1}) or (\ref{H2}) are,

\begin{eqnarray}
E(\{v_i\})	& = & u^2 + u\mathcal{C}_1  + \frac{\alpha}{\eta^2}\mathcal{C}_1^2 + \eta^{-2}\zeta^2 - (\omega_{1} + \omega_{2})^2 \nonumber \\
	& - & [u^2 - (\omega_{1} + \omega_{2})^2] \prod_{i=1}^{N}\frac{v_{i} - u -\eta}{v_{i} - u} \nonumber \\
	& - & \eta^{-2}\zeta^{2} \prod_{i=1}^{N}\frac{v_{i} - u +\eta}{v_{i} - u},
\end{eqnarray}
\noindent where the $\{v_i\}$ are solutions of the BAE (\ref{BAE1}) and $N$ is the total number of atoms. We can choose arbitrarily the spectral parameter $u$. In Fig. 2 we show the dimensionless ground state $E_0/\mu_1$ versus the relative external potential $\mu_2/\mu_1$ for different number of atoms $N$ and for a particular choice of the others parameters. 

\begin{figure}
\begin{center}
\includegraphics[scale=0.6]{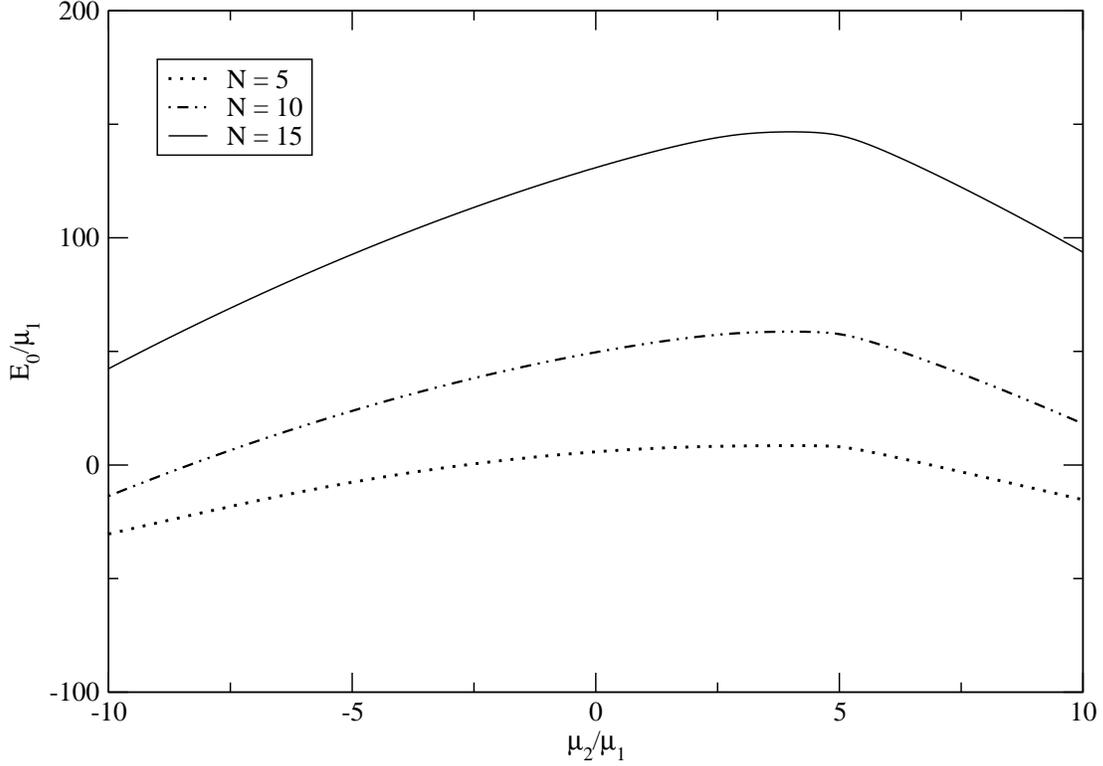}
\caption{In the figure we have the dimensionless ground state $E_0/\mu_1$ versus the relative external potential $\mu_2/\mu_1$   for different values of the total number of atoms $N$ and for the following choice of the parameters: $U_{aajj} = U_{bbjj}=U_{abjk} = 1$, $\;U_{aa12}=U_{bb12}=2$,  $\;\epsilon_{a1} = -\epsilon_{a2} = -2$, $\;\epsilon_{b1} = -\epsilon_{b2} = 1$, $\;\mu_1=1$, $\;\Omega_{jk} = 0.5,\;j,k=1,2$, $u=0$ and $\omega_1=\omega_2=\eta=\zeta = 1$. }
\end{center}
\label{f2}
\end{figure}

\subsection{The two-well model with $n$ on well states}

The Hamiltonian of the system is,

\begin{eqnarray}
H & = & \sum_{p=a,b}\;\;\sum_{j=1}^{n} U_{ppjj}N_{pj}N_{pj}  \nonumber \\ 
&+& \frac{1}{2}\sum_{p=a,b}\;\;\sum_{ j,k=1 (j\neq k)}^{n} U_{ppjk} N_{pj}N_{pk} + \sum_{j,k=1}^{n} U_{abjk} N_{aj}N_{bk} \nonumber \\  
& - &  \sum_{j=1}^{n} \mu_{j}(N_{aj} - N_{bj}) + \sum_{j=1}^{n} \epsilon_{aj}N_{aj} + \sum_{j=1}^{n} \epsilon_{bj} N_{bj} \nonumber \\
& - & \sum_{j,k=1}^{n} \Omega_{jk}(a_{j}^{\dagger}b_{k} + b_{k}^{\dagger}a_{j}).
\label{H3}
\end{eqnarray}
\noindent The parameters in this model are like the parameters in the Hamiltonian (\ref{H1}), we just remark that $U_{ppjk} = U_{ppkj}$. The Hamiltonian (\ref{H3}) describes $s$-wave scattering between the atoms in all $n$ on well states and the tunnelling of the atoms between the wells $a$ and $b$ to the same or to different on well states.

It is important to note that if we turn off the tunnelling,\;$\Omega_{jk}=0$, the Hamiltonian (\ref{H3}) describes two decoupled Bose-Einstein condensates with $n$ states in each one, $\{|n_{pj}\rangle\},\;p=a,b,\;j=1,\ldots,n,$  and only one pure vector state for each condensate, $|\psi_p\rangle = \bigotimes_{j=1}^n |n_{pj}\rangle,\;p=a,b$ , with the total vector state of the system the tensor product of those pure vectors states,

$$ |\Psi_T\rangle = |\psi_a\rangle\otimes|\psi_b\rangle.$$

 The energies, $E_a$ and $E_b$, are respectively,
 
\begin{eqnarray}
E_a & = & \sum_{j=1}^{n} U_{aajj}N_{aj}N_{aj} + \frac{1}{2}\sum_{ j,k=1 (j\neq k)}^{n} U_{aajk} N_{aj}N_{ak}  +   \sum_{j=1}^{n} (\epsilon_{aj} - \mu_{j})N_{aj},
\label{EaH3}
\end{eqnarray}

\begin{eqnarray}
E_b & = & \sum_{j=1}^{n} U_{bbjj}N_{bj}N_{bj} + \frac{1}{2}\sum_{ j,k=1 (j\neq k)}^{n} U_{bbjk} N_{bj}N_{bk} +  \sum_{j=1}^{n} (\epsilon_{bj} + \mu_{j})N_{bj}.
\label{EbH3}
\end{eqnarray}

The total energy is, $E = E_a + E_b$, and the ground state in each condensate depends only of the scattering interactions $U_{aaij}$ and $U_{bbij}$, and the external potentials $\mu_j$. We have $2n$ conserved quantities, $[H,I_{pj}]= 0,\;I_{pj} = N_{pj},\;j=1,\ldots,n,\;(p=a,b),$ and so the total number of atoms is also a conserved quantity, $N =  \sum_{p=a,b}\sum_{j=1}^n I_{pj}$.

When we turn on the tunnelling,\;$\Omega_{jk} \neq 0$,the BEC are now coupled by Josephson tunnelling and the total number of atoms,

$$N = N_{a1} + N_{a2}  + N_{b1} + N_{b2},$$

\noindent continue a conserved quantity, but now, $I_{pj}$, are no more conserved because of the tunnelling amplitudes,

$$[H,N] = 0,\qquad [H,I_{pj}] \neq 0.$$
\noindent  If $\Omega_{jk}=\Omega_{kj}=0,\; \forall\; j \neq k$, we have another conserved quantities, $[H,J_{j}]=0$, with $J_j = N_{aj} + N_{bj},\;j=1,\ldots,n,$ and so the total number of atoms is a conserved quantity, $N =  \sum_j^n J_j$.

 The state space is spanned by the base $\{|n_{a1},\ldots,n_{an},n_{b1},\ldots,n_{bn}\rangle\}$ and we can write each vector state as 

\begin{equation}
  |n_{a1},\ldots,n_{an},n_{b1},\ldots,n_{bn}\rangle = \frac{1}{\sqrt{\prod_{j=1}^n n_{aj}!n_{bj}!}}\prod_{j=1}^n (a_{j}^{\dagger})^{n_{aj}}(b_{j}^{\dagger})^{n_{bj}}|0\rangle,
\label{state1}   
\end{equation}

\noindent where $|0\rangle = |0_{a1},\ldots,0_{an},0_{b1},\ldots,0_{bn}\rangle$ is the vacuum vector state in the Fock space. We can use the states (\ref{state1}) to write the matrix representation of the Hamiltonian (\ref{H3}). The dimension of the space increase very fast when we increase $N$,

$$	d = \frac{(L -1 +N)!}{(L-1)!N!}, $$

\noindent where $L = 2n$ is the total number of states in both wells and $N$ as a constant $c$-number, $N = n_{a1} + n_{a2} + n_{b1} + n_{b2}$. In the case where we have only two states \cite{GSantosaa} (one in each well, $n=1$) the dimension is $d = N + 1$.

Now we use the co-multiplication property of the Lax operators to write,

\begin{equation}
  L(u) = L_{1}^{\Sigma a}(u + \sum_{j=1}^{n}\omega_{j})L_{2}^{\Sigma b}(u - \sum_{j=1}^{n}\omega_{j}).
\label{LH2}
\end{equation}
\noindent Following the monodromy matrix (\ref{monod}) we can write the operators,

\begin{eqnarray}
\pi(A(u)) &=& (uI+I\sum_{j=1}^n \omega_j + \eta\sum_{j=1}^n N_{aj})(uI  - I\sum_{j=1}^n \omega_j + \eta\sum_{j=1}^n N_{bj}) \nonumber \\
          &+& \sum_{j,k=1}^n s_jt_kb_j^{\dagger}a_k, \label{piAn}\\
\pi(B(u)) &=& (uI + I\sum_{j=1}^n \omega_j + \eta\sum_{j=1}^n N_{aj})\sum_{j=1}^n t_jb_j + \eta^{-1}\zeta\sum_{j=1}^n t_jb_j, \label{piBn}\\
\pi(C(u)) &=& (\sum_{j=1}^n s_ja_j^{\dagger})(uI  - I\sum_{j=1}^n \omega_j + \eta\sum_{j=1}^n N_{bj}) \nonumber \\
          &+& \eta^{-1}\zeta\sum_{j=1}^n t_jb_j^{\dagger}, \label{piCn}\\
\pi(D(u)) &=& \sum_{j,k=1}^n s_jt_ka_j^{\dagger}b_k + \eta^{-2}\zeta^2 I. \label{piDn}
\end{eqnarray}

Taking the trace of the operator (\ref{LH2}) we get the transfer matrix

\begin{eqnarray}
t(u) & = & u^2I + u\eta N + [\eta^{-2}\zeta^{2} - \sum_{j,k=1}^{n}\omega_{j}\omega_{k}]I  \nonumber\\
	 & + & \eta(\sum_{j=1}^{n}\omega_{j})\sum_{j=1}^{n}(N_{bj} - N_{aj}) + \eta^{2}\sum_{j,k=1}^{n}N_{aj}N_{bk} \nonumber \\
 	 & + & \sum_{j,k=1}^{n}s_{j}t_{k}(a_{j}^{\dagger}b_{k} + b_{k}^{\dagger}a_{j}).
\label{tu2} 
\end{eqnarray}

\noindent From (\ref{C14b}) we identify the conserved quantities of the transfer matrix (\ref{tu2}),

\begin{eqnarray}
\mathcal{C}_0 
	 & = &  [\eta^{-2}\zeta^{2} - \sum_{j,k=1}^{n}\omega_{j}\omega_{k}]I  \nonumber\\
	 & + & \eta(\sum_{j=1}^{n}\omega_{j})\sum_{j=1}^{n}(N_{bj} - N_{aj}) + \eta^{2}\sum_{j,k=1}^{n}N_{aj}N_{bk} \nonumber \\
 	 & + & \sum_{j,k=1}^{n} s_{j}t_{k}(a_{j}^{\dagger}b_{k} + b_{k}^{\dagger}a_{j}), 
\end{eqnarray}

\begin{eqnarray}
\mathcal{C}_1 &=&  \eta N,
\end{eqnarray}

\begin{eqnarray}
\mathcal{C}_2 &=& I.
\end{eqnarray}

The Hamiltonian (\ref{H3}) is related with the transfer matrix (\ref{tu2}) by the equation,

\begin{equation}
 H = u^2 I + u\mathcal{C}_1  + \frac{\alpha}{\eta^2}\mathcal{C}_1^2   + [\eta^{-2}\zeta^{2} - \sum_{j,k=1}^{n}\omega_{j}\omega_{k}]I - t(u),
\label{H4} 
\end{equation}
\noindent where we have the following identification between the parameters,

$$\alpha =  U_{aajj} = U_{bbjj}, \qquad 2\alpha = U_{aajk} = U_{bbjk} \; (j\neq k),$$

$$  2\alpha - \eta^2 = U_{abjk},$$

$$ \eta(u - \sum_{j=1}^{n}\omega_{j}) = \epsilon_{aj} - \mu_{j},\qquad  \eta(u + \sum_{j=1}^{n}\omega_{j}) = \epsilon_{bj} + \mu_{j} $$

$$ \Omega_{jk} = s_j t_k, \qquad \sum_{j=1}^{n} s_j t_j = \zeta, \qquad  j,k = 1,\ldots ,\; n. $$

We use as pseudo-vacuum the product state, 

$$|0\rangle = |\{0\}_{n}\rangle_{a}\otimes|\{0\}_{n}\rangle_{b},$$

\noindent with 
 
 $$|\{0\}_{n}\rangle_{p} = |0_{1},0_{2},\ldots,0_{n}\rangle_{p},$$ 

\noindent belonging to the direct sum space associated to the states and denoting the Fock vacuum state for the well $p$ $(p = a,b)$. For this pseudo-vacuum we can apply the algebraic Bethe ansatz method in order to find the Bethe ansatz equations (BAE),

\begin{eqnarray}
\frac{\eta^{2}(v^2_{i} - \sum_{j,k=1}^{n}\omega_{j}\omega_{k})}{\zeta^{2}} & = & \prod_{j \ne i}^{N}\frac{v_{i}-v_{j}-\eta}{v_{i}-v_{j}+\eta}, \;\;\;\;\;  i,j = 1,\ldots , N.
\label{BAE2}
\end{eqnarray}

The eigenvectors $\{ |v_1,v_2,\ldots,v_N\rangle \}$ of the Hamiltonian (\ref{H3}) or (\ref{H4}) and of the transfer matrix (\ref{tu2}) are

\begin{equation}
|\vec{v}\rangle \equiv  |v_1,v_2,\ldots,v_N\rangle = \prod_{i=1}^N \pi(C(v_i))|0 \rangle,
\end{equation}
\noindent and the eigenvalues of the Hamiltonian (\ref{H3}) or (\ref{H4}) are,

\begin{eqnarray}
E(\{ v_i \}) & = &   u^2 + u\mathcal{C}_1  + \frac{\alpha}{\eta^2}\mathcal{C}_1^2 + \eta^{-2}\zeta^{2} - \sum_{j,k=1}^{n}\omega_{j}\omega_{k} \nonumber \\
  & - &  (u^2 - \sum_{j,k=1}^{n}\omega_{j}\omega_{k}) \prod_{i=1}^{N}\frac{v_{i} - u -\eta}{v_{i} - u} \nonumber \\
  & - &  \eta^{-2}\zeta^{2}\prod_{i=1}^{N}\frac{v_{i} - u +\eta}{v_{i} - u}. 
\end{eqnarray}
\noindent where the $\{v_i\}$ are solutions of the BAE (\ref{BAE2}) and $N$ is the total number of atoms. We can choose arbitrarily the spectral parameter $u$.

\section{Summary}
We have introduced a new family of two-well models with arbitrary $n$ on well state in each well and derived the Bethe ansatz equations  and the corresponding eigenvalues. 
These models were obtained through a combination of Lax operators constructed using the Heisenberg-Weyl Lie algebra. 
An interesting aspect of these models is that no selection rules for tunnelling between the on well states are present, 
the atoms can thus tunnel to the same or different on well states. We believe that the models proposed, as they can furnish precise results on physical quantities as the energy gap, entanglement and ground state fidelity, have potential applications in
studies such as those involving quantum metrology  \cite{drummond, oberth, campbell} in the context of ultracold atoms. Also, the obtained Hamiltonians may be useful in the context of NMR techniques \cite{nmr1, oliv,oliv2} as an alternative to the usual nuclear quadrupole Hamiltonian.

\section*{Acknowledgments}
The authors acknowledge Capes/FAPERJ (Coordena\c{c}\~ao de Aperfei\c{c}oamento de Pessoal de N\'{\i}vel Superior/Funda\c{c}\~ao de Amparo \`a Pesquisa do Estado do Rio de Janeiro) and CNPq (Conselho Nacional de Desenvolvimento Cient\'{\i}fico e Tecnol\'{o}gico) for financial support.


\section*{References}
\begin{thebibliography}{10}

\bibitem{early} Cornell E. A.  and Wieman C. E., \textit{Rev. Mod. Phys.} \textbf{74} (2002) 875.


\bibitem{angly} Anglin J. R.  and  Ketterle W., \textit{Nature} \textbf{416} (2002) 211.

  
\bibitem{drummond} He Q. Y., Reid M. D., Vaughan T. G., Gross C., Oberthaler M. and Drummond P. D., \textit{Phys. Rev. Lett.} \textbf{106} (2011) 120405.


\bibitem{jon1} Zhou H.-Q., Links J., Gould M. and McKenzie R., \textit{J. Math. Phys.} \textbf{44} (2003) 4690.


\bibitem{jon2} Zhou H.-Q., Links J. and McKenzie R. H., \textit{Int. Jour. Mod. Phys. B} \textbf{17} (2003) 5819.

\bibitem{jonjpa} Links J., Zhou H.-Q., McKenzie R. H.  and Gould M. D., \textit{J. Phys. A} \textbf{36} (2003) R63.

\bibitem{key-3} Foerster A. , Links J. and Zhou H.-Q., \textit{Classical and quantum non-linear integrable systems: theory and applications}, Editor A. Kundu, IOP Publishing, Bristol and Philadelphia, (2003) 208.


\bibitem{dukelskyy} Dukelsky J., Dussel G., Esebbag C.  and Pittel S., \textit{Phys. Rev. Lett.} \textbf{93} (2004) 050403.

\bibitem{Ortiz} Ortiz G., Somma R., Dukelsky J. and Rombouls S., \textit{Nuclear Physics B} \textbf{707} (2005) 421.
 
\bibitem{Kundu} Kundu A., \textit{Theoretical and Mathematical Physics} \textbf{151} (2007) 831.

\bibitem{eric5} Foerster A.  and Ragoucy E., \textit{Nuclear Physics B} \textbf{777} (2007) 373.

\bibitem{GSantosaa} Links J., Foerster A., Tonel A. P. and Santos G., \textit{Ann. Henri Poincar\'e} \textbf{7} (2006) 1591.

\bibitem{GSantos} Santos G., Foerster A., Roditi I., Santos Z. V. T. and Tonel A. P., \textit{J. Phys. A: Math. Theor.} \textbf{41} (2008) 295003.


\bibitem{GSantos11} Santos G., \textit{J. Phys. A: Math. Theor.} \textbf{44} (2011) 345003.

\bibitem{hertier} H\'eritier M., \textit{Nature} \textbf{414} (2001) 31.

\bibitem{batchelor} Batchelor M. T., \textit{Physics Today} \textbf{60} (2007) 36.

\bibitem{fst} Faddeev L. D., Sklyanin E. K. and Takhtajan L. A., \textit{Theor. Math. Phys.} \textbf{40} (1979) 194.

\bibitem{ks} Kulish P. P. and Sklyanin E. K, \textit{Integrable Quantum Field Theories: 
Proceedings of the Symposium Held at Tvrminne, Finland - Lecture Notes in Physics} Editor: J. Hietarinta and C. Montonen, \textbf{151}, Springer-Verlag, Berlin, (1982) 61.

\bibitem{takhtajan} Takhtajan L. A., \textit{Quantum Groups: 
Proceedings of the 8th International Workshop on Mathematical Physics Held at the Arnold Sommerfeld Institute, Clausthal, FRG - 
 Lecture Notes in Physics}, Editor: H. -D. Doebner and J. -D. Hennig, \textbf{370}, Springer-Verlag, Berlin, (1990) 3.


\bibitem{korepin} Korepin V. E., Bogoliubov N. M.  and Izergin A. G., \textit{Quantum inverse scattering method and correlation functions}, Cambridge University Press, Cambridge, (1993).


\bibitem{faddeev} Faddeev L. D., \textit{Int. J. Mod. Phys. A } \textbf{10} (1995) 1845.


\bibitem{korepin1} Izergin A. G.  and Korepin V. E., \textit{Lett. Math. Phys.} \textbf{6} (1982) 283.

\bibitem{yang} Yang C.N., \textit{Phys. Rev. Lett.} \textbf{19} (1967) 1312.

\bibitem{korepin2} Izergin A. G.  and Korepin V. E., \textit{Nuc. Phys. B} \textbf{205} (1982) 401.

\bibitem{ek} Essler F. H. L.  and Korepin V. E., \textit{Exactly solvable models of strongly correlated electrons}, World Scientific, Singapore, (1994). 

\bibitem{ek2} Essler F. H. L., Frahm H., G\"ohmann F., Kl\"umper A. and Korepin V. E., \textit{The one-dimensional Hubbard Model}, Cambridge University Press, Cambridge, (2005). 

\bibitem{blz} Bazhanov V., Lukyanov S.  and Zamolodchikov A. B., \textit{Commun. Math. Phys.} \textbf{177} (1996) 381.

\bibitem{lipatov} Lipatov L., \textit{JETP Lett.} \textbf{59} (1994) 596.

\bibitem{korch} Faddeev L., Korchemsky G., \textit{Phys. Lett. B} \textbf{342} (1995) 311.

\bibitem{belitsky} Belitsky A.V., Braun V.M.,  Gorsky A.S., Korchemsky G.P., \textit{Int. J. Mod. Phys.} \textbf{A 19} (2004)  4715.

\bibitem{jimbo85} Jimbo M., \textit{Lett. Math. Phys.} \textbf{10} (1985) 63. 

\bibitem{jimbo86} Jimbo M., \textit{Field Theory, Quantum Gravity and Strings: 
Proceedings of a Seminar Series Held at DAPHE, Observatoire de Meudon, and LPTHE, Universit\'e Pierre et Marie Curie, Paris - Lecture Notes in Physics}, Editor: H. J. de Vega and N. S\'anchez, \textbf{246}, Springer-Verlag, Berlin, (1986) 335.

\bibitem{drinfeld} Drinfeld V. G., \textit{Quantum groups: Proc. Int. Congress of Mathematicians}, Editor: A. M. Gleason, Providence, RI: American Mathematical Society, (1986) 798.

\bibitem{frt} Reshetikhin N. Yu, Takhtajan L. A. and Faddeev L. D., \textit{Leningrad Math. J.} \textbf{1} (1990) 193.

\bibitem{faddeev2} Faddeev L. D., \textit{40 Years in Mathematical Physics - World Scientific Series in 20th Century Mathematics}, \textbf{2}, World Scientific Publishing Co. Pte. Ltd., Singapore, (1995).

\bibitem{dorey}  Dorey N., \textit{J. Phys. A: Math. Theor.} \textbf{42} (2009) 254001.

\bibitem{batchelor2} Guan X.-W. , Batchelor M. T. and Lee C., \textit{Fermi gases in one dimension: From Bethe Ansatz to experiments}, arXiv:1301.6446 [cond-mat.quant-gas]. 


\bibitem{kino1} Kinoshita T.,  Wenger T. and  Weiss D.S., \textit{Science} \textbf{305} (2004) 1125.

\bibitem{kino2} Kinoshita T.,  Wenger T. and  Weiss D.S., \textit{Nature} \textbf{440} (2006) 900.

\bibitem{kitagawa} Kitagawa T., Pielawa S., Imambekov A., Schmiedmayer J., Gritsev V., and Demler E., \textit{Phys. Rev. Lett.} \textbf{104} (2010) 255302.

\bibitem{haller} Haller E., Gustavsson M., Mark M.J., Danzl J.G., Hart H., Pupillo G., N\"agerl H.-C., \textit{Science} \textbf{325} (2009) 1224.

\bibitem{liao} Liao Y., Rittner C., Paprotta T., Li W., Partridge G.B., Hulet R.G., Baur S.K. and Mueller E.J., \textit{Nature} \textbf{467} (2010) 567.

\bibitem{coldea} Coldea R., Tennant D.A., Wheeler E.M., Wawrzinska E., Prabhakaran D., Telling M. Habicht K.
Smeibidil P., Kiefer K., \textit{Science} \textbf{327} (2010) 177.

\bibitem{nmr1} Fel'dman, E. B., Pyrkov, A. N., Zenchuk, A. I., \textit{Philosophical Transactions of The Royal Society A} \textbf{370} (2012) 4690.




\bibitem{Kuznet} Kuznetsov V.B. and Tsiganov A.V., \textit{J. Phys. A: Math. Gen.} \textbf{22} (1989) L73.

\bibitem{QYHEDrumm} He Q.Y., Vaughan T. G., Drummond P. D., and  Reid M. D., \textit{New J. Phys. } \textbf{14} (2012) 093012.

\bibitem{kuhnert} Kuhnert M., Geiger R. , Langen T., Gring M., Rauer B., Kitagawa T., Demler E., Adu Smith D., and Schmiedmayer J., \textit{Multimode dynamics and emergence of a characteristic length-scale in a one-dimensional quantum system}, \textit{Phys. Rev. Lett.} \textbf{110} (2013) 090405.

\bibitem{rubeni} Rubeni D., Foerster A., Mattei E., and Roditi I., \textit{Nuclear Physics B} \textbf{856} (2012) 698.

\bibitem{oliv} Oliveira I.S., Bonagamba T.J., Sarthour R., Freitas J.C.C., and deAzevedo E.R., \textit{NMR Quantum Information Processing}, Elsevier, Amsterdam, (2007).

\bibitem{oliv2} Araujo-Ferreira A. G., Auccaise R., Sarthour R. S., Oliveira I. S., Bonagamba T. J. and Roditi I., \textit{Classical bifurcation in a quadrupolar NMR system}, \textit{Phys. Rev. A} \textbf{87} (2013) 053605. 


\bibitem{Roditi} Roditi I., \textit{Brazilian Journal of Physics} \textbf{30} (2000) 357.  

\bibitem{oberth} C. Gross, T. Zibold, E. Nicklas, J. Est\`eve and M. K. Oberthaler, \textit{Nature} \textbf{464} (2010) 1165.


\bibitem{campbell} Hennig, H., Witthaut, D. and Campbell, D.~K., \textit{Physical Review A} \textbf{86} (2012) 051604(R).


\end{thebibliography}
\end{document}